\documentstyle{mn}
\input{epsf}

\begin{document}

\title{A search for counter-rotating stars in S0 galaxies}

\author[K. Kuijken, D. Fisher \& M.R. Merrifield]{
Konrad Kuijken$^1$\thanks{
Visiting Scientist, Dept. of Theoretical Physics,
University of the Basque Country, Bilbao, Spain}, David Fisher$^1$ and 
Michael R. Merrifield$^2$\\
   $^1$ Kapteyn Astronomical Institute, University of Groningen, 
P.O. Box 800, 9700 AV Groningen, The Netherlands\\
   $^2$ Department of Physics, University of Southampton, 
Highfield SO9 5NH, UK
}

                         
\maketitle

\begin{abstract}
We have obtained high signal-to-noise spectra along the major axes of
28 S0 galaxies in order to search for the presence of disk stars on
retrograde orbits.  Full line-of-sight velocity distributions were
extracted from the data, and the velocity distributions were modelled
as arising from the superposition of populations of stars on prograde
and retrograde orbits.  We find no new cases in which a significant
fraction of disk stars lie on retrograde orbits; an identical analysis of
NGC~4550 does reveal the previously-known counter-rotating stellar disk
in this system. Upper limits determined for each object indicate that no more
than $\sim 5\%$ of the observed disk star light could arise from
counter-rotating stellar components.  These results suggest that
previously-discovered disk galaxies with counter-rotating stars are
exceptional and that (at 95\% confidence) at most $10\%$ of S0
galaxies contain significant counter-rotating populations. The most
likely value for the fraction of such S0 galaxies lies closer to 1\%.
This result contrasts with the prevalence of counter-rotating gas in
these systems; combining our new observations with existing data, we
find that $24 \pm 8\%$ (1-$\sigma$ error) of the gas disks in S0
galaxies counter-rotate relative to their stellar components.
\end{abstract}

\begin{keywords}
line: profiles -- galaxies: elliptical and lenticular, cD -- galaxies:
           kinematics and dynamics -- galaxies: structure
\end{keywords}

\section{INTRODUCTION}
Over recent years, a number of techniques have been developed which
use high signal-to-noise ratio spectra of galaxies to obtain detailed
information as to their stellar kinematics (eg Bender 1990; Rix and
White 1992; van der Marel and Franx 1993; Gerhard 1993; Kuijken and
Merrifield 1993; Saha \& Williams 1994; Statler 1995).  These
techniques, in combination with photometric measurements, have
provided new insights into the properties of a variety of galaxy
components, including nuclear disks (van der Marel et al.~1994),
the outskirts of elliptical galaxies (Bender, Saglia \& Gerhard 1994;
Carollo et al.~1995), peanut-shaped bulges (Kuijken \& Merrifield
1995), and bars (Merrifield \& Kuijken 1995). 

However, perhaps the most dramatic discovery has been the detection of
disk galaxies in which a significant fraction of the stars are on
retrograde orbits.  Such a counter-rotating population was first
discovered by Rubin et al.~(1992) in the Virgo lenticular
NGC~4550. This system was found to contain a disk in which half of the
stars circulate in one direction while the other half orbit in the
opposite direction (Rix et al.\ 1992).  The only other documented
example of a large-scale counter-rotating stellar disk is in the Sab
galaxy NGC~7217, in which Merrifield \& Kuijken (1994) discovered that
the disk is made up from two distinct components with approximately
two-thirds on prograde orbits and one-third on retrograde orbits.
This smaller retrograde population only showed up after a careful
analysis of the line-of-sight velocity distribution (LOSVD) derived
from absorption-line spectra at a number of radii in the system: disk
stars in the main prograde component follow approximately circular
orbits, and so produce large peaks in the observed LOSVDs at the
projected circular speed of the galaxy (relative to its systemic
velocity); the counter-rotating population shows up in the complete
velocity distributions as secondary peaks at close to minus the
projected circular velocity.

One possibly-related phenomenon is the relatively high frequency with
which gas disks in early-type disk galaxies are observed to
counter-rotate.  Bertola et al.\ (1992) find that approximately 20\%
of the gas disks found in S0 galaxies rotate in the opposite direction
to the stars in these systems.  There is no {\it a priori} reason to
suppose that stars cannot form in such gas just as they do in gas on
more normal prograde orbits, and so we might expect counter-rotating
stellar populations to form in these systems.  However, the gas disks
in these early-type galaxies are usually restricted to their cores,
and so any counter-rotating stars that they produce are likely to be
in a similar small region.  Support for this scenario has come from
recent observations of the Sa galaxy NGC~3593, which has been found to
contain both a counter-rotating gas disk in its core and a
counter-rotating stellar disk on a similar small spatial scale
(Bertola et al.\ 1996).  Similarly, Prada et al.\ (1996) have found
that there is a counter-rotating stellar component in the bulge region
of the Sb galaxy NGC~7331.  Although star formation in central
counter-rotating gas disks could result in these relatively
small-scale phenomena, it could not produce the large-scale
counter-rotating stellar populations that we see in NGC~4550 and
NGC~7217, and it is not clear how closely these two phenomena are
related.

One problem in trying to interpret large-scale stellar
counter-rotation is that only a very small number of disk galaxies
have been subjected to the sort of detailed kinematic analysis
required to detect modest fractions of stars on retrograde orbits.
The possibility therefore exists that such counter-rotating
populations could lie undetected in a significant percentage of all
disk galaxies.  Further, it is difficult to make a statistical
interpretation of the existing data: most of the known cases of
counter-rotation (both stellar and gaseous) were detected
serendipitously, and this rather heterogeneous data set lacks
well-defined selection criteria.

We have therefore obtained high signal-to-noise ratio spectra for a
well-defined sample of 28 early-type disk galaxies.  In this paper, we
present the kinematic analysis of these data, and use them to
calculate the frequency of both stellar and gaseous counter-rotation
in such systems.

\section{DATA REDUCTION AND ANALYSIS}
The galaxies chosen for this study (see Table~1) were selected from
the Third Reference Catalogue of Bright Galaxies (de Vaucouleurs et
al.~1991). They belong to Hubble morphological types S0, SB0, and SA0,
which hereafter will be collectively referred to as S0 galaxies. The
sample consists of objects in a variety of environments, with members
drawn from the field, small groups and rich clusters.  We restricted
our sample of disk galaxies to early-types because their smooth light
profiles and comparatively uniform stellar populations enable accurate
measures of the stellar kinematics. This selection also has the
benefit that the possible complicating influences on measurement and
interpretation from gas, dust, and star forming regions are minimized
compared to later-type spirals. Furthermore, it has been argued that
the frequency of counter-rotating stars might be higher in
earlier-type disk galaxies (Merrifield \& Kuijken 1994).  In order to
maximize the projection into the line of sight of the stellar
rotational motion (and hence maximize the characteristic signal from
any counter-rotating population), preference was given to observing
edge-on objects. Apart from NGC~3998 (inclination 35$^\circ$), all
galaxies in our sample have inclinations of at least 50$^\circ$, with
more than half inclined more than 80$^\circ$. Of the sample galaxies,
only NGC~4550 was known to exhibit counterrotating stars when this
project was started.

\begin{table}
\caption{Parameters for the galaxy sample observed}
 \begin{tabular} {lrlrrc} 
\multicolumn{1}{c}{Galaxy} & \multicolumn{1}{c}{$B^\circ_T$}  & Type &
PA$_{\rm maj}$ & Obs & Gas\\ \hline
               &       &             &      &   &    \\ 
NGC$\;\;\;$128 & 12.66 & S0 pec      &   1  & M & $-$\\ 
NGC 1184       & 13.44 & S0/a        & 167  & M & $+$\\ 
NGC 1332       & 11.25 & S0$^-$      & 148  & M &    \\ 
NGC 1461       & 12.66 & SA0$^\circ$ & 155  & L &    \\ 
NGC 1611       & 12.95 & SB0$^+$     &  77  & M & $+$\\ 
NGC 2549       & 12.04 & SA0$^\circ$ & 177  & M &    \\ 
NGC 2560       & 14.08 & S0/a        &  93  & L &    \\ 
NGC 2612       & 13.5  & S0$^-$      & 115  & M & $-$\\ 
NGC 3098       & 13.00 & S0          &  90  & M & $+$\\ 
NGC 3115       &  9.75 & S0$^-$      &  43  & L &    \\ 
NGC 3384       & 10.75 & SB0$^-$     &  53  & L &    \\ 
NGC 3412       & 11.35 & SB0$^\circ$ & 155  & L &    \\ 
NGC 3607       & 10.87 & SA0$^\circ$ & 120  & L & $+$\\ 
NGC 3941       & 11.27 & SB0$^\circ$ &  10  & L & $-$\\ 
NGC 3998       & 11.54 & SA0$^\circ$ & 140  & L & $+$\\
NGC 4026       & 11.60 & S0          & 178  & L & $+$\\ 
NGC 4036       & 11.52 & S0$^-$      &  85  & L & $+$\\ 
NGC 4111       & 11.62 & SA0$^+$     & 150  & L,M & $+$\\ 
NGC 4124       & 12.12 & SA0$^+$     & 114  & M & $+$\\ 
NGC 4179       & 11.97 & S0          & 143  & M & $+$\\ 
NGC 4251       & 11.54 & SB0         & 100  & L &    \\ 
NGC 4350       & 11.87 & SA0         &  28  & L & $+$\\ 
NGC 4550$\dagger$ & 12.4  & SB0      & 178 & L &    \\
NGC 4710       & 11.84 & SA0$^+$     &  27  & M &    \\ 
NGC 4754       & 11.38 & SB0$^-$     &  23  & L &    \\ 
NGC 4762       & 11.05 & SB0$^\circ$ &  32  & L & $+$\\ 
NGC 5308       & 12.43 & S0$^-$      &  60  & M &    \\ 
NGC 5866       & 10.87 & SA0$^+$     & 128  & L & $+$\\ 
NGC 7332       & 11.96 & S0          & 155  & M & $-$\\ 
\hline
 \end{tabular}

Total $B$ band `face-on' magnitudes, $B^\circ_T$, corrected for
Galactic and internal extinction, and for redshift from RC3 (de
Vaucouleurs et al.~1991).  For NGC 1184 we give $m_B$ and for NGC 1611
we give $m_{\rm FIR}$ both from the RC3. For NGC 2612 we quote the
value given in the NASA Extragalactic Database.  Galaxy types are
as given in the RC3. Major axis position angles, PA$_{\rm maj}$, are
taken from the RC3 except for NGC 1184 which was determined by eye.
The ``Obs'' column denotes whether the galaxy was observed from the
Lick (L) or the Multiple Mirror Telescope Observatory (M). The final column
indicates whether gas was detected, and, if so, whether it follows
prograde ($+$) or retrograde ($-$) orbits. $\dagger$NGC~4550 was known
to contain counterrotating disks before this survey was started, and
is not included in the final statistics.

\end{table}

The spectra analyzed here were obtained with the Shane Telescope
through the KAST spectrograph at Lick Observatory, and the Multiple
Mirror Telescope at Mount Hopkins, using the Red Channel
Spectrograph. The Lick observations were taken through a $2\ {\rm
arcsec} \times 2.6\ {\rm arcmin}$ slit with a $1200\ {\rm line}\ {\rm
mm}^{-1}$ grating, resulting in a spectral resolution of 3.1 \AA\
(FWHM) corresponding to a instrumental dispersion of $\sigma$=75
km/s. The MMT spectrograph was configured with a $1200\ {\rm line}\
{\rm mm}^{-1}$ grating and a $1.25\ {\rm arcsec} \times 3\ {\rm
arcmin}$ slit giving a resolution of 2.6 \AA\ (instrumental
$\sigma$=63 km/s). For both sets of observations, the spectral region
around the Mg b triplet at 5170 \AA\ was observed with integration
times of typically 40 -- 60 min.

The techniques employed here to determine the galaxy line profiles are
the expansion of the LOSVD into a truncated Gauss-Hermite series (van
der Marel and Franx 1993) and the unresolved Gaussian decomposition
(UGD) method of Kuijken and Merrifield (1993). In the first method the
velocity distribution is parametrised by the Gauss-Hermite series of
orthogonal functions yielding two measures of the deviations of the
line profile from a gaussian: a parameter $h_3$ which measures
asymmetric deviations, and a parameter $h_4$ measuring symmetric
deviations. The UGD method extracts the stellar LOSVD by modeling the
velocity distribution produced by the Doppler broadening and velocity
shift in the galaxy spectrum. The line profile is modeled as the sum
of a set of unresolved (ie overlapping) gaussians with fixed means and
dispersions. The optimal LOSVD is obtained by convolving the modeled
LOSVD with spectra of template stars and minimizing the least-squares
comparison with the galaxy spectrum by varying the amplitudes of the
gaussian components.

\section{RESULTS}

In this section we present the LOSVDs derived using the UGD method for
our sample.

\begin{figure}
\epsfxsize=\hsize \epsfbox{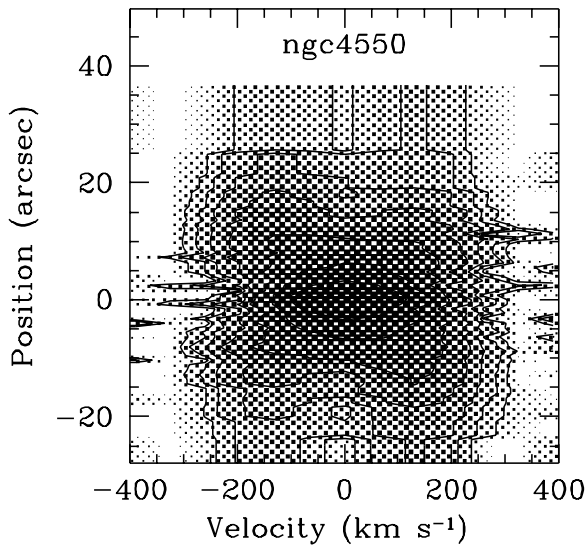}
\caption{The line-of-sight velocity distribution for NGC~4550, as
derived with the unresolved gaussian decomposition method. The
greyscales and contours indicate the projected phase space density
(stars arcsec$^{-1}$ km s$^{-1}$) on the major axis. Contours are
logarithmically spaced, separated by factors of $2^{1/2}$. In
agreement with the results of Rix et al.~(1992), two counterrotating
stellar components are clearly identifiable.}
\label{fig-ngc4550}
\end{figure}

As a guide to the eye, we first show the result for the galaxy NGC
4550 in which the signature of counter-rotating disks is clearly seen
as a splitting of the line profile into two approximately equal
strength peaks (Fig.~\ref{fig-ngc4550}).  This figure demonstrates
that our data and analysis technique are indeed capable of revealing
counter-rotating components, in agreement with the double-gaussian
analysis of this galaxy's spectra by Rix et al.~(1992).

Figure~\ref{fig-losvds} displays the stellar LOSVDs for the remainder
of our sample of objects as derived from the UGD method. It is
apparent from Fig. 1 that NGC~4550 is unique among the galaxies in our
sample as none displays a comparable LOSVD. A number of the objects
(e.g., NGC~3384, NGC~4111, NGC~4762) have line profiles suggestive of
independent kinematic components such as nuclear disks and bars.
However, there are no obvious signs of counter-rotation.

\begin{figure*}
\epsfxsize=16truecm 
\caption{Velocity distributions along the major axes 
of the sample galaxies. Positions and radial velocities are given with
respect to the centre of each galaxy. The contour levels are separated
by factors of $2^{1/2}$ in density; greyscale levels jump every other
contour. Features at the lowest levels in these panels may be
artefacts of the particular stellar template spectrum chosen for the
analysis (see text).}
\label{fig-losvds}
\end{figure*}
\begin{figure*}
\epsfxsize=16truecm 
\contcaption{}
\end{figure*}

\begin{figure}
\epsfxsize\hsize\epsfbox{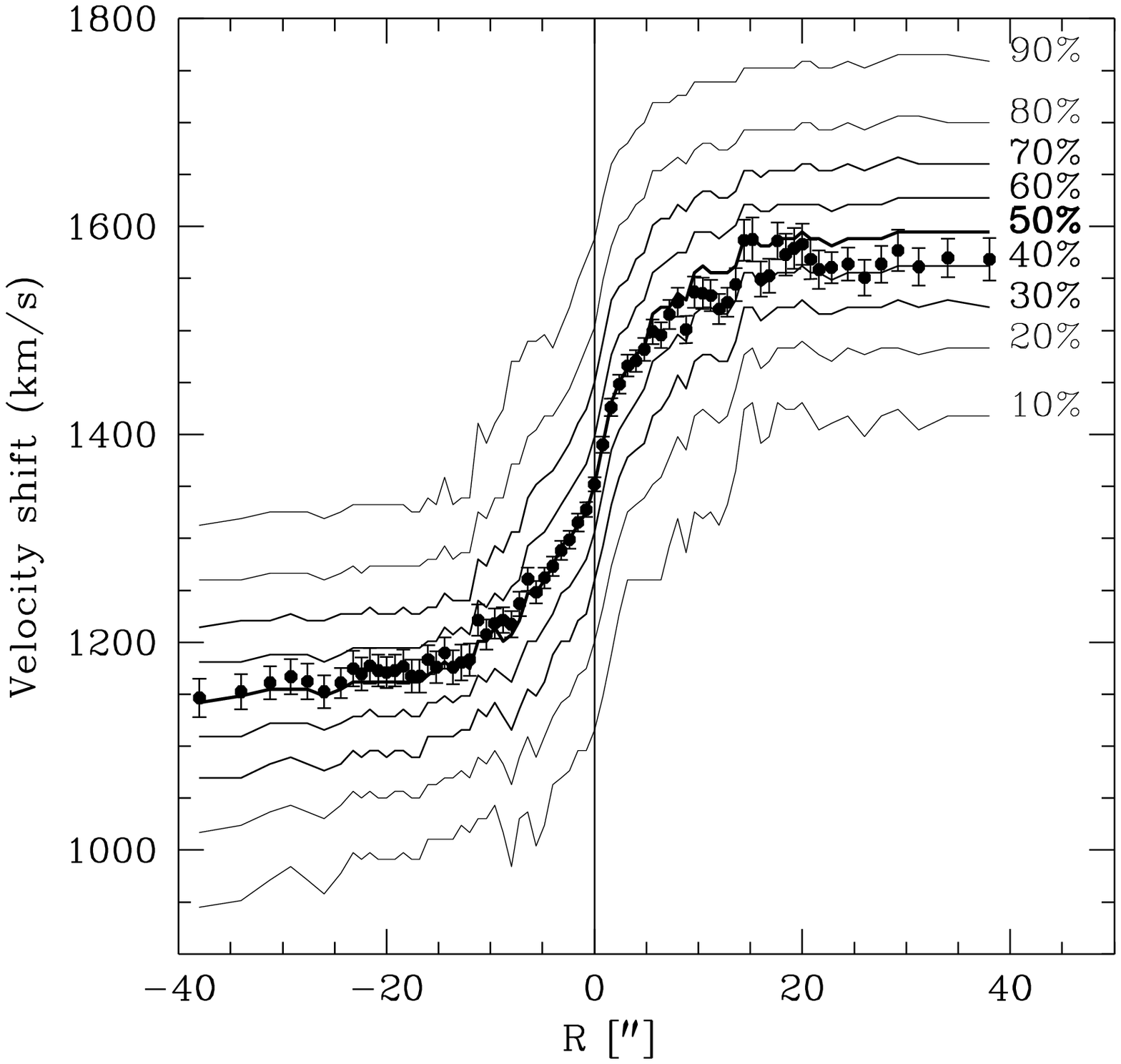}
\epsfxsize\hsize\epsfbox{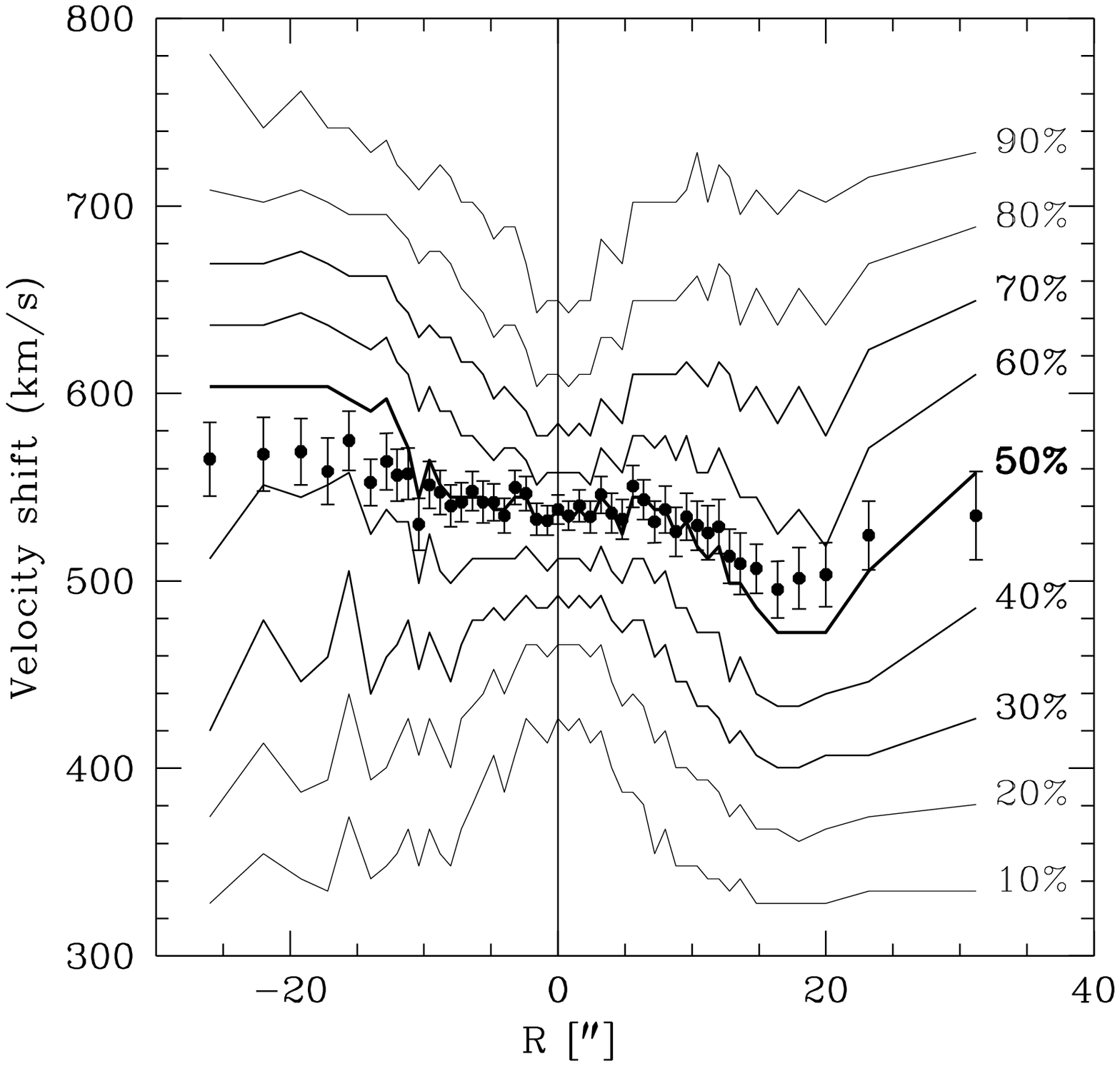}
\caption{Contours of the cumulative velocity distributions for NGC
4036 (top) and NGC~4550 (bottom). (10\% of the stars have radial
velocity below the line labelled 10\%, etc.) This figure illustrates
the strong limits that our data can place on the fraction of
counterrotating stars. Points with error bars give the mean velocity
at each radius.}
\label{fig-cumlosvd}
\end{figure}

The line profiles have been analyzed further by generating cumulative
velocity distributions which indicate the fraction of stars at
successive velocity increments. An example is shown in
Fig.~\ref{fig-cumlosvd} for the major axis of NGC~4036. In this
representative case, no peculiarities are present in the `rotation
curve' with the only distinguishing feature being a low-speed tail
which reaches its greatest skewness at a radius of 10 arcseconds.
This produces an asymmetry in the line profile resulting in $h_3$
terms of magnitude 0.1 as output by the Gauss-Hermite expansion. Such
tails are generically expected in disks with radially decreasing
density and velocity dispersions (Kuijken \& Tremaine 1992);
line-of-sight projection through the edge-on galaxy further enhances
the skewness. For comparison, we also plot the cumulative velocity
distribution for NGC~4550, which shows that roughly equal numbers of
stars populate each component.

\subsection{Upper limits on counter-rotating populations}

A straightforward procedure for estimating the upper limit on the
counter-rotating stars is as follows. Starting from the measured
LOSVDs (which we will call the prograde model), we construct a
retrograde model by inverting all velocities with respect to the
galaxy's systemic velocity.  We then convolve both the prograde and
retrograde velocity distributions with the stellar template spectrum,
resulting in two model spectra: one for the prograde stars, one for
the retrograde ones. We then attempt to model the observed galaxy
spectrum as a linear superposition of the pro- and retrograde models,
and evaluate the statistical bounds on the coefficient of the
retrograde model. Only those spectra which cover the flat part of the
rotation curve, outside the bulge, are included in the analysis. The
upper limits obtained in this way range between 2 and 6\%
(3-$\sigma$). 

\begin{figure}
\epsfxsize\hsize\epsfbox{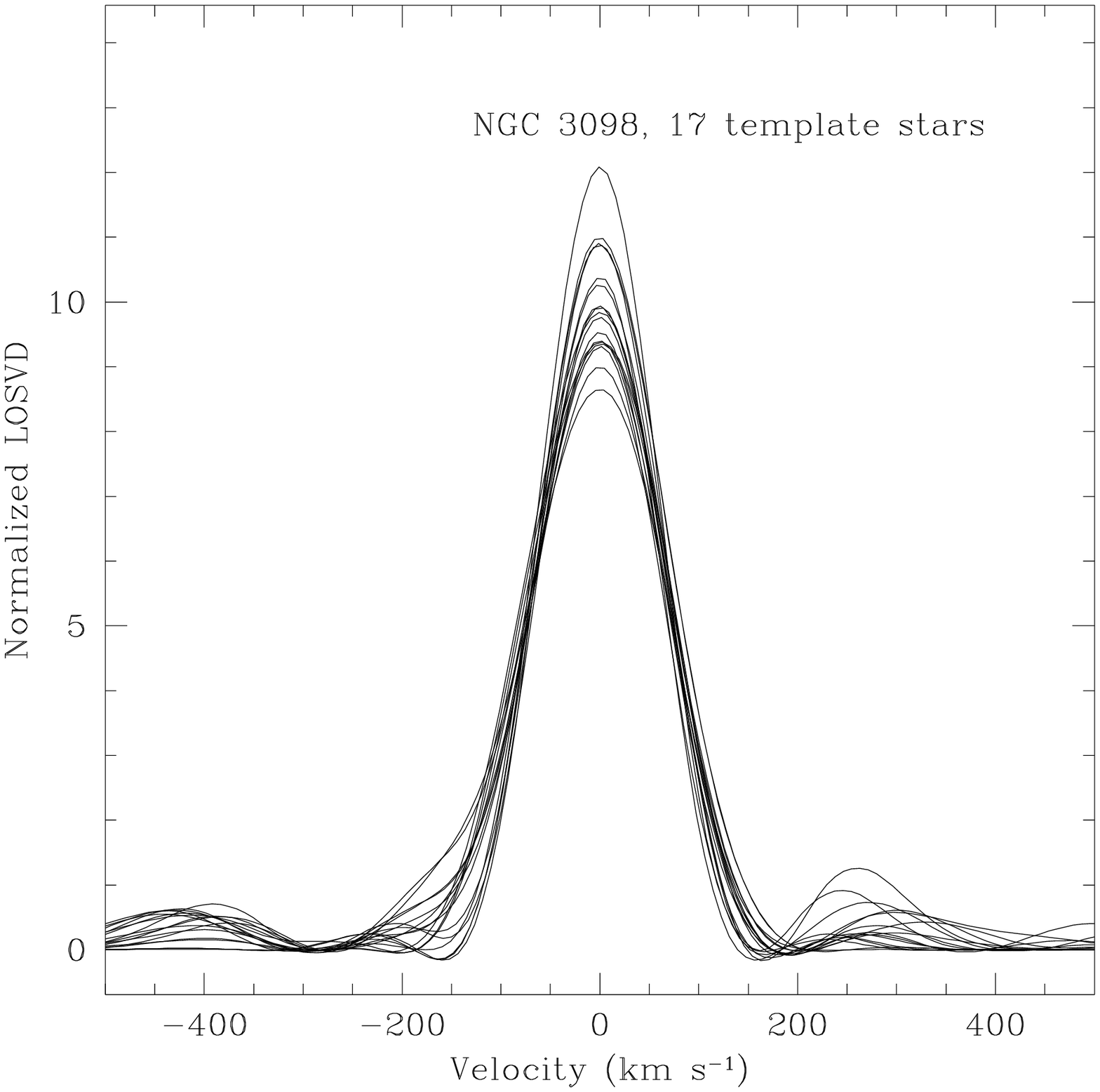}
\epsfxsize\hsize\epsfbox{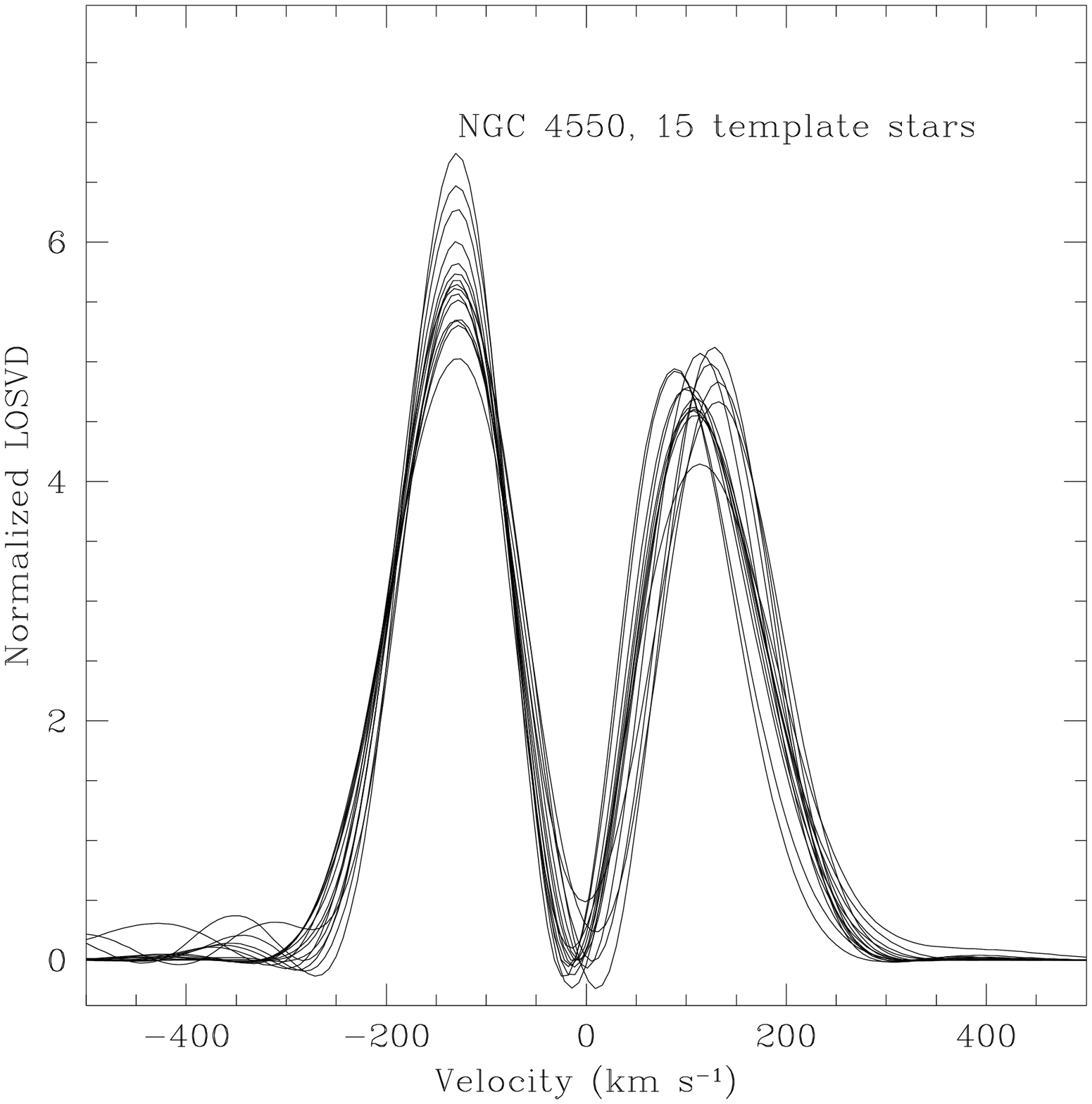}
\caption{LOSVDs derived from a point in the disks of NGC~3098 (top)
and NGC~4550 (bottom).  The distributions have been derived using
stellar templates with a range of spectral (F -- M), and it is
clear that the choice of template type affects the derived velocity
distribution at around the 5\% level.}
\label{fig-mismatch}
\end{figure}

These formal errors reflect pure photon statistics, without allowance
for possible systematic errors. The most important such effect is
template mismatch, which occurs when the stellar template chosen
differs from the average stellar spectrum in the galaxy.  In order to
investigate the importance of this effect, we have derived the LOSVDs
for a number of the observed galaxies using template stars with a
range of spectral types.  As Fig.~\ref{fig-mismatch} illustrates, this
analysis reveals that template mismatch can produce spurious features
in the derived distributions at around the 5\% level.  It is this
effect which is probably responsible for the three galaxies (NGC~2612,
NGC~3098, NGC~5866) where the LOSVDs shown in Fig.~\ref{fig-losvds}
are mildly suggestive of counter-rotation.  Template mismatch thus
makes it impossible to constrain the presence of such counter-rotating
stars to a level better than 5\% from the present analysis.  It should
also be remembered that the UGD analysis, which does not allow
negative line profiles, contains a slight bias towards creating
positive LOSVD peaks at the 1-$\sigma$ level at velocities where there
is in reality no flux.

Limits on any counterrotating disk stars in the bulge regions of the
galaxies are more difficult to obtain. In these regions the background
from the kinematically hot bulge is more difficult to separate from
any superimposed disk, and the distinction between disk and bulge is
not clear without further information about vertical kinematic
gradients. We therefore do not emphasize the bulge regions in this
study; at first sight, none of our sample galaxies show strong
indications for counter-rotating central stellar disks.

\subsection{The fraction of S0 galaxies with 
counter-rotating stellar disks}

Given that we find no evidence for counter-rotation in our sample of
28 objects (excluding NGC~4550, which was not observed as part of our
unbiased survey), what limit does our result place on the fraction $p$
of all galaxies which can contain counter-rotating populations?
Simple binomial distribution statistics show that if $p$ were greater
than 10\%, the probability of not finding even one case in our sample
would be less than 5\%. The 95\% confidence upper limit for $p$ is
thus 10\%.  \footnote{It is unwise to incorporate NGC~4550 itself into these
statistics, since we observed it precisely because it had been shown
to contain a counter-rotating disk.  However, including it with full
weight in the statistics would yield a somewhat biassed confidence
interval of $3.5^{+13.5}_{-3.4}$\%, not dramatically different from
the upper limit quoted above.}

\subsection{The fraction of S0 galaxies with counter-rotating gaseous disks}

Our spectral observations of these galaxies cover the 4959\AA\ and
5007\AA\ lines of [O{\sc iii}], as well as H$\beta$ at 4861\AA.  We
can thus also use the data to search for emission lines in these
systems, and measure the fraction in which any gas that we detect is
found to be counter-rotating.  Since the sample galaxies were not
selected on the basis of their gaseous properties, these data provide
an unbiassed estimate for the fraction of S0 galaxies containing
counter-rotating gas disks.  The last column in Table~1 shows in which
galaxies emission line gas was detected, and in which systems it
follows retrograde orbits.  Of the 28 galaxies in the sample, gas was
detected in 17.  Thirteen of these systems are prograde, while the
remaining four contain gas on retrograde orbits.  We thus find that
the counter-rotating gas disks make up $24 \pm 10\%$ (1-$\sigma$
error; the 90\% confidence interval is 8 -- 46\%).  This value agrees
very well with the figure of $\sim 20\%$ that Bertola et al.\ (1992)
obtain from a sample of 15 S0 galaxies selected on the basis of their
known extended gas emission.  The two samples have three galaxies in
common,\footnote{For one of the galaxies in both samples, NGC~128, we
find that the gas is on retrograde orbits, in disagreement with
Bertola et al.'s tabulated result.  Our result is confirmed by
Emsellem et al.\ (1995)} and of the combined sample of 29 galaxies in
which gas is detected, we find that the orbits are retrograde in 7
systems.  Our best estimate for the fraction of S0 galaxy gas disks
which counter-rotate relative to the stellar component is thus
$24 \pm 8\%$ (1-$\sigma$ errors; 12 -- 40\% with 90\% confidence).

\section{DISCUSSION}
The goal of the present study has been to search for the presence of
counter-rotating stars in S0 galaxies through an analysis of their
line-of-sight velocity distributions. No new cases of stellar
counter-rotation have been found.  The upper limits derived from this
analysis imply that no more than $\sim 5\%$ of the disk stars in our
sample galaxies could remain undetected on retrograde orbits.  This
analysis implies that the previously-discovered examples of disk
galaxies with counter-rotating stars are quite rare: at 95\%
confidence we can say that no more than 10\% of S0 galaxies contain
detectable counterrotating stellar disks.  Successful theories of
galaxy formation must, however, still take account for these unusual
objects, and so it is worth discussing how these systems might have
formed.

Galaxy mergers provide the most obvious mechanism by which
counter-rotating populations might form.  However, mergers of
equal-mass disk galaxies with antiparallel spins result in systems
that would not be identified as a disk galaxy -- the heating of the
initial galaxies is too great to preserve the disk structure (Barnes
1992).  Less dramatic merging of satellites on to galaxy disks can
produce counter-rotation (Thakar \& Ryden 1996), but still
cannot explain the equal-mass counter-rotating disks seen in NGC~4550.

A more general scenario for the formation of galaxies with
counter-rotating disks is via the steady infall of gas onto a
preexisting disk system (see Bettoni \& Galletta 1992; Rix et al.\
1992; Merrifield \& Kuijken 1994). Numerical simulations by Thakar \&
Ryden (1996) indicate that the accretion of diffuse gas can, indeed,
lead to the formation of large-scale counter-rotating components.  

The prevalence of counter-rotating gaseous disks that we find in S0
galaxies provides support for a scenario of this kind.  However, most
of these gas disks are quite small, and cannot be responsible for
large-scale counter-rotating stellar disks such as NGC~4550 or
NGC~7217, but are more likely to produce centrally-concentrated
retrograde stellar populations such as the one discovered in NGC~3593
(Bertola et al.\ 1996).  The two galaxies in our sample which contain
more extensive counter-rotating gas disks -- NGC~3941 and NGC~7332
(Fisher 1994; Fisher, Illingworth \& Franx 1994) -- show no signs of
corresponding counter-rotating stellar components.  In the case of
NGC~7332, the emission-line gas is clearly not yet relaxed, so the
lack of associated stars is perhaps not a surprise. In NGC~3941 the
gas disk appears well settled, with a radius of about 30 arcseconds
(2.7kpc). Roberts et al.~(1991) quote a total HI gas mass of $1.3
\times 10^9M_\odot$ for this galaxy ($H_0=50\rm km\,s^{-1}kpc^{-1}$),
which, if it all resides in the retrograde gas disk, would be
responsible for about 5\% of the mass in the inner 2.7 kpc -- more if
there is also a substantial molecular component.  The absence of a
counter-rotating stellar component in this system then suggests that
star-formation may be inhibited in gas on retrograde orbits, possibly
because of interaction with the stellar winds from the prograde stars.
However, high-resolution mapping of this system in HI and CO is
required before any firm conclusion can be drawn.  If it does prove to
be difficult to produce stars in massive retrograde gas disks, then
the mystery surrounding the few rare cases which do contain
large-scale counter-rotating stellar populations deepens still
further.

\section*{ACKNOWLEDGEMENTS}
The data presented in this paper were obtained at Lick Observatory
which is operated by the University of California and with the
Multiple Mirror Telescope which is a joint facility of the Smithsonian
Institute and the University of Arizona.  Much of the analysis was
performed using {\sc iraf}, which is distributed by NOAO.  MRM is
supported by a PPARC Advanced Fellowship (B/94/AF/1840).  We thank
Marijn Franx for a careful reading of the manuscript, and the referee,
Gerry Gilmore, for constructive comments.

\end{document}